\documentclass[twocolumn,pra,showpacs,aps]{revtex4}
\usepackage{amssymb}
\usepackage{subeqn}
\usepackage{epsfig}
\usepackage{dcolumn}
\usepackage{bm}

\newcommand{\ui}{\mathrm{i}}
\newcommand{\be}[0]{\begin{equation}}
\newcommand{\ee}[0]{\end{equation}}
\newcommand{\bea}[0]{\setlength\arraycolsep{2pt}\begin{eqnarray}}
\newcommand{\eea}[0]{\end{eqnarray}}
\newcommand{\bse}{\begin{subequations}}
\newcommand{\ese}{\end{subequations}}
\newcommand{\ra}{\rangle}

\begin{document}

\title{Center-of-mass interpretation for bipartite purity analysis of $N$-party entanglement}
\author{Miguel A. Alonso$^{1,2}$}
\author{Xiao-Feng Qian$^{1,2,3}$}
\author{J.H. Eberly$^{1,2,3}$}
\affiliation{$^{1}$The Institute of Optics, University of Rochester,
Rochester, NY 14627, USA\\
$^{2}$Center for Coherence and Quantum Optics, University
of Rochester, Rochester, New York 14627, USA\\
$^{3}$Department of Physics \& Astronomy, University
of Rochester, Rochester, New York 14627, USA
}
\date{\today }

\pacs{03.65.Ud, 03.65.Yz, 42.50.-p}

\begin{abstract}
We provide a graphical description of the entanglement of pure-state multiparty systems based on an analogy between a bipartite purity analysis and the centroid of a collection of point masses. This description applies to quantum systems with $N$ parties, each with an arbitrary number of (discrete) states. The case of $N$ qubits is highlighted for simplicity. This geometric description illustrates some of the restrictions in the form of inequalities that apply to entanglement in multiparty systems. 
\end{abstract}
\maketitle

\noindent{\bf Introduction.} In tandem with widespread experimental efforts to create entanglement in many-party systems, an ongoing theoretical effort has been aimed at quantifying many-party entanglement. The entanglement existing collectively between all parties of an $N$-party system, called 
genuinely multipartite entanglement (GME), is of particular importance since it plays a central role in many applications. 
Entanglement 
is best identified via its opposite, biseparability. A pure state $|\Psi\ra$ is biseparable if it can be written as $|\Psi\ra = |\psi_{\rm A}\ra \otimes |\psi_{\rm B}\ra$, where $|\psi_{\rm A}\ra$ and $|\psi_{\rm B}\ra$ are pure states. A mixed state is biseparable if it can be written as a sum of pure separable states in any bipartition; otherwise, the state is 
genuinely multipartite 
entangled \cite{Horodecki-etal-09, Huber-etal-10}. 
Quantifying 
GME
has proved to be a challenging task. Previous studies have produced witnesses and/or lower bounds (see \cite{SMHR-etal-13}). 
Areas of open $N$-party entanglement issues include multi-electron atomic ionization \cite{Becker-etal-12}, multilevel coding for quantum key distribution \cite{Multi-QKD} and multiparty teleportation \cite{MultiParty-tele, MultiEnt-tele}. 

In order to gain further insight into entanglement, new techniques must be considered, especially if they offer useful physical and/or geometrical analogies. Several geometrically-inspired treatments have been proposed, including descriptions by way of Pauli operator expectation values \cite{Liang}, and Bloch sphere state representations \cite{Regula}, amongst others.   
In this letter we approach entanglement in multiparty systems
through its connection with each party's purity. We show that this leads to a geometric representation arising from a direct link to point-particle center-of-mass theorems. This new step has valuable visualization advantages, carrying insights from few-party examples to arbitrary $N$-party entanglements.

\noindent{\bf Measures of purity.} For a pure $N$-party system, the entanglement of one party with the remaining $N-1$ parties determines the purity of that party's quantum state when the rest of the system is traced out. 
In fact, when a system has a quantum state that is mixed, it is so because of its entanglement with parties not  explicitly considered \cite{Qian-Eberly-10, Haroche}. 
%
The purities of each party following the tracing out of the remaining ones can then serve as the basis for characterizing entanglement in multiparty systems. 
We follow this approach by using a particular measure of purity for the individual parties that we show has an intuitive geometric interpretation, and lends itself for the description not only of each party but also of the complete system. We start by describing this measure for an individual party, and then we extend these ideas to the full system.

Consider an $M$-state single party, which  
(following the tracing out of the other parties) 
is described by a 
$M\times M$ density matrix $\rho$. 
The standard definition of purity for this party is ${\rm tr}(\rho^2)$. Its inverse, the Schmidt weight \cite{K}, is an entanglement monotone that 
gives a measure of the effective number of significant eigenvalues the matrix has. The purity takes values in the interval $[1/M,1]$, so the Schmidt weight is between 1 and $M$. These measures are invariant to local unitary transformations. 
Several other measures of purity have been defined that are monotonic functions of ${\rm tr}(\rho^2)$. In particular, we use 
\begin{equation}
Q=\sqrt{\frac{M{\rm tr}(\rho^2)-1}{M-1}}.
\end{equation}
This measure has the desirable property of varying between 0 and 1, with 0 corresponding to a maximally mixed state and 1 to a pure state. 
More importantly, its geometric interpretation described below makes it useful for visualizing entanglement between many parties. It is worth mentioning that 
both the Schmidt weight and
this measure are also used in the classical study of polarization of light and other vector wave phenomena, where instead of states one has two \cite{BornAndWolf} or three \cite{Qian-Eberly-12,Samson,Friberg,Colin2} Cartesian field components. 

\noindent{\bf Geometric interpretation for one party.} We now propose a geometric construction for interpreting $Q$ for an individual party in terms of a simple mechanical analogy. As discussed later, this construction provides insight into the characterization of entanglement in multiparty systems and the relations that constrain it. 
By using an appropriate local unitary transformation, the party's density matrix can be diagonalized:
\begin{equation}
\bar{\rho}=\mathbf{U}^{\dagger}\rho\mathbf{U}={\rm diag}(\lambda_1,\lambda_2,...,\lambda_M),
\end{equation}
where, without loss of generality, we order the states so that $\lambda_1\ge\lambda_2\ge\ldots\ge\lambda_M\ge0$, with $\sum_{m=1}^M\lambda_m=1$. We henceforth refer to this diagonalized representation as the {\it Schmidt representation} \cite{Schmidt,Fedorov-Miklin-14}, and denote it with an overbar. Since ${\rm tr}(\rho^2)={\rm tr}(\bar{\rho}^2)$, we can write
\begin{equation}
Q=\sqrt{\frac{M\sum_{m=1}^M\lambda_m^2-1}{M-1}}.
\end{equation}
Clearly, $Q=1$ holds only when $\lambda_1=1$ and $\lambda_{m>1}=0$, while $Q=0$ is true only when all eigenvalues are equal, $\lambda_m=1/M$.

\begin{figure}
\begin{center}
\includegraphics[scale=0.5]{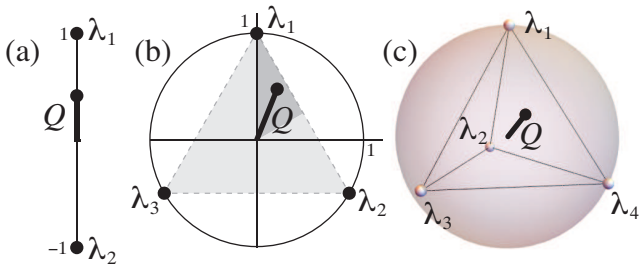}
\caption{Definition of $Q$ as the distance between the origin and the center of mass of $M$ point masses of magnitudes $\lambda_m$ at unit distances to the origin and mutually equidistant, within an Euclidean space of dimension $M-1$. (a) For $M=2$ the space is a line, which also contains the center of mass. (b) For $M=3$ the space is a plane, and the center of mass is constrained to the interior of a triangle (gray), whose corners are the three masses. In fact, since $\lambda_1 \ge \lambda_2 \ge \lambda_3 \ge 0$, the center of mass is within the darker region. (c) For $M=4$, the space is a volume, and the center of mass is constrained to the interior of a tetrahedron whose corners are the four masses.
}
\end{center}
\end{figure}

The geometric construction is the following: Consider a Euclidean space of dimension $M-1$ and imagine a set of $M$ point masses in this space, all at a unit distance from the origin (and hence over a unit hypersphere), and each equidistant to all the others, the distance being $\sqrt{2M/(M-1)}$. Let the magnitudes of these masses be the eigenvalues $\lambda_m$. The measure $Q$ is then the distance between the center of the sphere (the origin) and the center of mass of the system. 
%
This idea is illustrated in Fig. 1 for the simplest cases of $M=2$, 3, and 4. 
For $M=2$, shown in Fig.~1(a), the two masses are at the points $\pm1$ along a line, and $Q=\lambda_1-\lambda_2$. 
(This case is unique in that $Q$ is linear in the eigenvalues.) 
%
For $M=3$, the three masses are equidistantly distributed along a unit circle, at the corners of an equilateral triangle, as shown in Fig.~1(b). 
For $M=4$ shown in Fig.~1(c), the four masses are over the surface of a unit sphere, at the corners of a regular tetrahedron. For $M\ge5$, they are at the surface of a unit hypersphere, at the corners of a regular simplex inscribed in this hypersphere.\\

\noindent{\bf Geometric interpretation for multiple parties.} We now discuss how this center-of-mass picture can be used to characterize entanglement in multiparty systems. For simplicity we start by assuming that all parties are qubits ($M=2$), and show that the geometric interpretation proposed allows us to understand the limitations in entanglement in such systems. Following the standard practice when describing qubits, we label the two states of each party not by integers from 1 to $M=2$, but by 0 and 1. A general pure state consisting of $N$ qubit parties has a wave function that can be written as
\begin{equation}
|\psi\rangle=\sum_{i_1,...,i_N=0,1} c_{i_1,...,i_N}|i_1,...,i_N\rangle
=\sum_{\bf i} c_{\bf i}|{\bf i}\rangle,
\end{equation}
where $c_{i_1,...,i_N}$ are complex coefficients normalized to unity. In the second step we introduced the shorthand ${\bf i}=i_1,...,i_N$, and $\sum_{\bf i}$ to indicate the sum for all qubits over the two values $0$ and $1$. While the global state is pure, the description of a specific party $n$ is in terms of a $2\times2$ density matrix $\rho_n$ resulting from tracing out all parties but the $n$th one: 
\begin{equation}
(\rho_n)_{j,k}=\sum_{{\bf i}\ne i_n}c_{{\bf i}|_{i_n=j}} ^*\,c_{{\bf i}|_{i_n=k}},\label{rho}
\end{equation}
where $\sum_{{\bf i}\ne i_n}$ indicates summation over all indices except $i_n$. 
We choose a Schmidt representation for all parties, so that the density matrices are diagonal:
\begin{equation}
(\bar{\rho}_n)_{0,1}=(\bar{\rho}_n)_{1,0}^*=\sum_{{\bf i}\ne i_n}\bar{c}_{{\bf i}|_{i_n=0}} ^*\,\bar{c}_{{\bf i}|_{i_n=1}}=0.\label{perpendicular}
\end{equation}
The measures $Q_n$ for each party can be written as
\begin{equation}
Q_n = \sqrt{2{\rm tr}(\bar{\rho}_n^2)-1}=\sum_{\bf i}(-1)^{i_n} \left|\bar{c}_{\bf i}\right|^2.\label{Qt}
\end{equation}
Equation~(\ref{Qt}) suggests an $N$-dimensional space in which the vector ${\bf Q}=(Q_1,...,Q_N)$ is defined, as well as a geometric interpretation for this vector. Consider an $N$-dimensional hypercube of side 2, centered at the origin. Let a collection of point masses be placed at the points with coordinates $[(-1)^{i_1},...,(-1)^{i_N}]$, {\it i.e.}, the corners of the hypercube, where the magnitude of each mass is the modulus squared of the corresponding coefficient, $\left|\bar{c}_{\bf i}\right|^2 $. The center of mass of all these point masses (given that their sum is unity) is then ${\bf Q}$. This is illustrated in Fig.~2(a) for $N=2$. For $N=3$ the point masses would be the corners of a cube of side 2 centered at the origin.
\begin{figure}
\begin{center}
\includegraphics[scale=0.5]{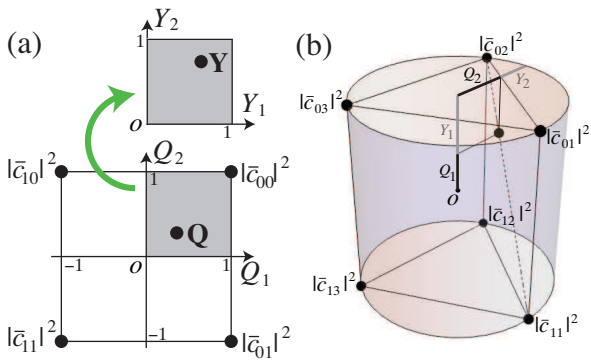}
\caption{(a) For $N$ qubits, the $N$-vector ${\bf Q}$ corresponds to the center of mass of a set of $2^N$ point masses of magnitude $|\bar{c}_i|^2$, placed at the corners of a hypercube of side 2 centered at the origin. For the $N=2$ case shown here, the hypercube reduces to a square.  (b) The same construction for a two-party system composed of a qubit and a qutrit, whose interpretation requires three dimensions. In both cases, the entanglement measures $Y_n=1-Q_n$ are also shown.}
\end{center}
\end{figure}

\noindent{\bf Entanglement vector and restrictions.} 
In general, measures of entanglement are restricted by inequalities referred to as monogamy relations. This includes relations limiting measures of entanglement such as the so-called tangle \cite{Coffman-etal-00,Osborne,Bai,Regulaetal,Etschka}, or related to Bell non-locality \cite{Toner,Su}. While not equivalent to those just mentioned, relations of this type also hold for the measures of entanglement discussed here, whose geometric interpretation we discuss in what follows. For the case of qubits, Higuchi {\it et al.} \cite{Higuchi} derived a form of these inequalities in terms of the eigenvalues of the density matrices for the bipartitions, and Walter {\it et al.} \cite{Walter} proposed a graphic representation in terms of polytopes over the space of the eigenvalues. For completeness, we present in the Supplemental Material \cite{Suppl.Matl.} a concise proof of these inequalities in terms of ${\bf Q}$, which take the form
\begin{equation}
N-2+Q_n\ge\sum_{n'\ne n}Q_{n'},\,\,\,\,n\in[1,N].\label{ineq}
\end{equation}
(An alternative, more geometric proof for the case $N=3$ is discussed in the next section.) 
These restrictions mean that not all the hypercube is accessible to ${\bf Q}$. They take a particularly simple form if we define a measure of entanglement of party $n$ with the rest of the parties as
\begin{equation}
Y_n = 1-Q_n,
\end{equation}
where $Y_n = 0$ indicates that party $n$ is completely separable from the rest, while $Y_n = 1$ indicates complete entanglement with the remaining parties. For qubits, the measure $Y_n $ is a valid entanglement measure, since it is simply twice the entanglement monotone $E_2(\rho_n)$ given in \cite{W}, which determines the conversion between different entanglement-valued states with certainty under LOCC \cite{Nielsen}. Also, it is easy to show that the von Neumann entropy is a monotonically increasing function of $Y_n$, {\it i.e.}, 
\setlength\arraycolsep{2pt}\begin{eqnarray}
S_n&=&-{\rm tr}(\rho_n\log_2\rho_n)\nonumber\\
&=&1-\frac{(2-Y_n)\log_2(2-Y_n)+Y_n\log_2(Y_n)}2,
\end{eqnarray}
Not only is this expression monotonic, but the limiting values $S_n=0,1$ correspond exactly to $Y_n=0,1$. It remains to be shown whether for $M\ge3$ the measure $Y_n$ remains monotonic under LOCC, although it is easy to show that the limits $S_n=0,1$ still correspond exactly to $Y_n=0,1$.

\begin{figure}
\begin{center}
\includegraphics[scale=0.5]{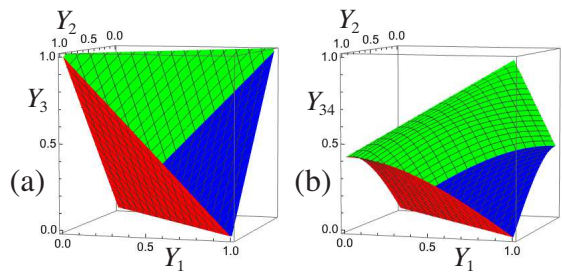}
\caption{Boundaries between the allowed and forbidden regions for three-party systems, dictated by (a) relation (\ref{ineqforY}) for three qubits, and (b) relations (\ref{twotwo4}) for two qubits and a qutetrit. The allowed regions are those on the side of the concavity formed by the three surfaces. The movie associated with (b) shows how these bounds are respected for one million randomly generated states.}\label{bounds}
\end{center}
\end{figure}

The restrictions in (\ref{ineq}) reduce to a simple polygon inequality when written in terms of $Y_n$ \cite{Qian-etal-15}:
\begin{equation}
Y_n \leq \sum_{n'\neq n}Y_{n'}.\label{ineqforY}
\end{equation}
Relations~(\ref{ineqforY}) can be visualized by defining an $N$-dimensional {\it entanglement vector} ${\bf Y}$ whose components are $Y_n$. As shown in Fig.~2(a), the space occupied by this vector is just a flipped version of the space occupied by ${\bf Q}$. The restrictions in (\ref{ineqforY}) mean that the $N$-dimensional hypervolume inhabitable by ${\bf Y}$ is not the unit hypercube but a simplex of hypervolume $1-1/(N-1)!$. For example, for $N=1$, only the point $Y_1=0$ is inhabitable out of the whole unit line segment, while for $N=2$, ${\bf Y}$ must be along the diagonal $Y_1=Y_2$ joining the points of minimal and maximal entanglement, $(0,0)$ and $(1,1)$. For $N=3$ the vector ${\bf Y}$ resides within a unit cube, but relation~(\ref{ineqforY}) implies that only half of the cube's volume is accessible, the bounds being shown in Fig.~\ref{bounds}(a).

The inequalities in (\ref{ineqforY}) state that the entanglement of one party with the rest cannot be more than the sum of the entanglements of the remaining parties. One can think of an analogy with classical shared resources, say, real estate ownership: the parties are property owners, and $Y_n$ represents the total joint property of party $n$ with other parties (where each party of a jointly owned property owns equal parts, see Supplemental Material \cite{Suppl.Matl.}). Clearly, party $n$ cannot have more joint property with others than all the others have with party $n$ and with each other. This inequality among the measures $Y_n$ is tight for qubits, meaning that the equality is achievable. 

\begin{figure}
\begin{center}
\includegraphics[scale=0.8]{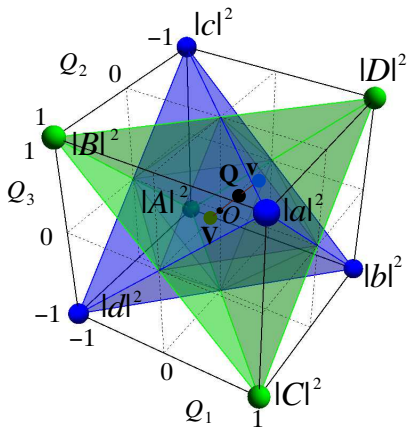}
\caption{The partial centers of mass ${\bf v}$ and ${\bf V}$ are each constrained to a tetrahedral volume whose corners are the masses in question. Since ${\bf v}$ and ${\bf V}$ are collinear with the origin, ${\bf Q}$ is contained in the union of the tetrahedra.}\label{tetra}
\end{center}
\end{figure}

\noindent{\bf Example: three-qubit pure state.} To gain geometric insight into the constraints in (\ref{ineqforY}), consider the case of three entangled qubits. For simplicity, we rename the coefficients $\bar{c}_{\bf i}$ according to
\setlength\arraycolsep{2pt}\begin{eqnarray}
\label{psiprime}
|\psi\rangle&=&a|000\rangle+C|001\rangle+D|010\rangle+b|011\rangle\nonumber\\
&+&B|100\rangle+d|101\rangle+c|110\rangle+A|111\rangle,
\end{eqnarray}
where lowercase (uppercase) letters are used for terms whose indices add up to an even (odd) number. These coefficients are assumed to satisfy the Schmidt representation conditions (\ref{perpendicular}). 
${\bf Q}$ corresponds to the center of mass of eight point masses at the corners of a cube of side 2 centered at the origin, as shown in Fig.~\ref{tetra}. It can be calculated in terms of two partial centers of mass: ${\bf v}$, corresponding to masses $|a|^2,|b|^2,|c|^2,|d|^2$ (blue), and ${\bf V}$, corresponding to masses $|A|^2,|B|^2,|C|^2,|D|^2$ (green). Each of these partial centers of mass is constrained to a tetrahedron (blue and green) whose corners are the masses in question. In principle, the global center of mass ${\bf Q}$ (necessarily along the line segment joining ${\bf v}$ and ${\bf V}$) could be anywhere in the cube (although the ordering convention for the eigenvalues means that the center of mass is within the positive octant). However, it is shown \cite{Suppl.Matl.} that the Schmidt representation conditions (\ref{perpendicular}) imply that the two partial centers of mass are collinear with the origin. That is, ${\bf v}$ and ${\bf V}$ are parallel, and so is then ${\bf Q}$, which is then constrained to the union of the two tetrahedra. The exclusion of ${\bf Q}$ from the regions not occupied by the two tetrahedra is equivalent to the constraint (\ref{ineqforY}), given the relation between ${\bf Q}$ and ${\bf Y}$. Given that, for qubits, $Q_n$ is just a linear combination (the difference) of the eigenvalues for the density matrix of party $n$, the colored region in the positive octant of Fig.~\ref{tetra} is a scaled version of the allowed region given in \cite{Walter} for the three-qubit case. 

{\bf Relations for parties with more than two states.} The proof of the restrictions in (\ref{ineqforY}) is only for parties with two states (qubits). Numerical tests suggest that these restrictions also hold for parties with more states, as long as the number of states of all parties is the same. If the different parties do not have the same number of states, these inequalities change, as we now discuss. 

Consider first the simplest such case, a qubit and a qutrit, shown in Fig.~2(b). This case is trivial, since in the Schmidt representation only two of the six coefficients can differ from zero, so the center of mass is constrained to the (dashed) line joining the corresponding two point masses. However, it illustrates the validity of the construction for general states and the insights that this geometric description gives: the center of mass is restricted by geometry to the interior of the simplex (in this case, a triangular prism) whose corners are the point masses, but the Schmidt representation imposes extra constraints in the combination of values that the masses can take, further reducing the region inhabitable by the center of mass (in this case, to a line). The boundaries of the region inhabitable by the center of mass are flat, but when one or more of the parties have three or more states, the space in which the entanglement vector is defined is of smaller dimensionality than the one where the center-of-mass interpretation holds because $Y_n=1-Q_n$ only uses the radial distance of the center of mass in the corresponding subspace. This reduction in dimensionality causes the corresponding boundaries of the region inhabitable by ${\bf Y}$ to be curved.  For example, for the qubit-qutrit case in Fig.~2(b), the center of mass is restricted to a straight line within the three-dimensional space, while the entanglement vector ${\bf Y}=(Y_1,Y_2)$ lives in a two-dimensional space and is restricted to the curve $[(1 - Y_1)^2 + 1]/2 = [2 (1 - Y_2)^2 + 1]/3$. 

Let's now consider a more complicated case with three parties: two qubits and a qutetrit (four states). This case can be thought of as the result of starting with four qubits and then merging qubits 3 and 4 into a qutetrit ``party 34''. 
It is shown in the Supplemental Material \cite{Suppl.Matl.} through the center-of-mass picture that the tight inequalities 
for these three parties are
\bse
\setlength\arraycolsep{2pt}\begin{eqnarray}
&|Y_1-Y_2|\le1-\sqrt{\frac{H[3(1-Y_{34})^2-1]}2},\\
&Y_{34}\le1 - \sqrt{\frac{(1 - Y_1)^2 + (1 - Y_2)^2 + H^2(1 - Y_1 - Y_2)}3},
\end{eqnarray}
\label{twotwo4}\ese
where $H(x) = {\rm Max}(0,x)$. The resulting inhabitable region allowed by relations~(\ref{twotwo4}) is shown in Fig.~\ref{bounds}(b), and it has curved boundaries for the reasons discussed earlier. The volume of this region is only $0.3457$, while for the case of three qubits shown in Fig.~\ref{bounds}(a) the volume allowed by relations~(\ref{ineqforY}) is $0.5$.

{\bf Bi-partitions of more than one party.} The inequalities in~(\ref{twotwo4}) can also be interpreted as giving a more complete picture of the entanglement constraints of a four qubit state. In addition to the four measures $Y_1,Y_2,Y_3,Y_4$, corresponding to bipartition of each party versus the rest, and restricted by relations~(\ref{ineqforY}), one can construct the measures $Y_{12}=Y_{34},\,Y_{13}=Y_{24},\,Y_{14}=Y_{23}$, corresponding to bipartitions of two versus two parties. This second set of measures is restricted by relations~(\ref{twotwo4}) and their analogs. It is shown in the Supplemental Material \cite{Suppl.Matl.} that these inequalities disagree only slightly with the corresponding relations for shared classical resources (which do have flat boundaries); the difference in the regions allowed by the quantum and classical relations is of the order of 10\% of the regions allowed by either. 


\noindent{\bf Concluding remarks.} 
We proposed an analytic
approach that we believe to be new and that exploits familiar mechanical intuition allowing new results to appear. Viewed as coming from a step in an unexplored direction, while centered on such a conventional measure as multiparty system purities, the results of this first step are attractive. A key to this geometric description is the Schmidt representation, which ensures that the density matrix is diagonal for any party following tracing out of the others. The modulus squared of each coefficient can then be thought of as the magnitude of a point mass in a space whose dimensionality is the sum of all the states for all parties minus the number of parties. The coordinates of the center of mass of all these point masses provides a description of the entanglement amongst the different parties.


\begin{acknowledgments}
We acknowledge financial
support from the National Science Foundation through awards
PHY-1507278, PHY-1068325, and PHY-1505189. We thank Rodrigo Guti\'errez Cuevas for useful discussions.
\end{acknowledgments}

\section{Supplemental Material}

\subsection{Derivation of constraints to the center of mass imposed by the Schmidt representation for the case of $N$ qubits}
Here we show how the constraints imposed by the Schmidt representation:
\be
(\bar{\rho}_n)_{0,1}=(\bar{\rho}_n)_{1,0}^*=\sum_{{\bf i}\ne i_n}\bar{c}_{{\bf i}|_{i_n=0}} ^*\,\bar{c}_{{\bf i}|_{i_n=1}}=0,\label{perpendicular}
\ee
restrict the coordinates of the center of mass for an $N$-qubit system according to the relation
\be
N-2+Q_n\ge\sum_{n'\ne n}Q_{n'},\,\,\,\,n\in[1,N],\label{ineq}
\ee
where
\be
Q_n = \sqrt{2{\rm tr}(\bar{\rho}_n^2)-1}=\sum_{\bf i}(-1)^{i_n} \left|\bar{c}_{\bf i}\right|^2.\label{Qt}
\ee

For simplicity, consider $n=1$, the other cases following by symmetry. Relation~(\ref{ineq}) can then be expressed as $I_1\ge0$ where
\bea
I_1&=&\frac12\left(N-2+Q_1-\sum_{n=2}^NQ_{n}\right)\nonumber\\
&=&\frac12\sum_{\bf i}\left[N-2+(-1)^{i_1}-\sum_{n=2}^N(-1)^{i_{n}}\right]|c_{\bf i}|^2\nonumber\\
&=&\sum_{{\bf i}'\ne i_1}\left[|\bar{c}_{0,{\bf i}'}|^2+\sigma_{{\bf i}'}\left(|\bar{c}_{0,{\bf i}'}|^2+|\bar{c}_{1,{\bf i}'}|^2\right)\right],\label{I1}
\eea
where the factor of $1/2$ was included for future convenience, Eq.~(\ref{Qt}) and the wavefunction's normalization were used in the second step, and
\be
\sigma_{{\bf i}'}=\frac12\left[N-3-\sum_{n=2}^N(-1)^{i_{n}'}\right].
\label{sig}
\ee

We assume (without loss of generality) a Schmidt representation where $Q_n\ge0$ for all $n\in[1,N]$, {\it i.e.},
\be
\sum_{i\ne i_n}|\bar{c}_{i|i_n=0}^{}|^2\ge\sum_{i\ne i_n}|\bar{c}_{i|i_n=1}^{}|^2.
\label{chore}
\ee
It is convenient to introduce the two vectors ${\bf \Psi}^{(j)}=(\bar{c}_{j,0,...,0},\bar{c}_{j,0,...,1},...,\bar{c}_{j,1,...,1})$ of dimensionality $\mu=2^{N-1}$, for $j=0,1$. The wavefunction's normalization means that $\left|{\bf \Psi}^{(0)}\right|^2+\left|{\bf \Psi}^{(1)}\right|^2=1$ and, from Eq.~(\ref{chore}),
\be
\left|{\bf \Psi}^{(0)}\right|\ge\left|{\bf \Psi}^{(1)}\right|.\label{2ge1}
\ee
By also introducing ${\bf w}=(\sigma_{0,...,0},\sigma_{0,...,1},...,\sigma_{1,...,1})$ with elements $w_m$, we can write Eq.~(\ref{I1}) as
\bea
I_1&=&\left|{\bf \Psi}^{(0)}\right|^2+\sum_{m=1}^{\mu}w_m\left[\left|\Psi_m^{(0)}\right|^2+\left|\Psi_m^{(1)}\right|^2\right]\nonumber\\
&\ge&\left|{\bf \Psi}^{(0)}\right|^2-\left|\Psi_1^{(0)}\right|^2-\left|\Psi_1^{(1)}\right|^2,\label{stepw}
\eea
where in the second step we used the fact that the term with $m=1$ is the only negative contribution to the sum, since $w_1=\sigma_{0,...,0}=-1$ and $w_{m\ne1}\ge0$. Hence, the proof reduces to showing that the last expression is not negative. For this purpose, let us rewrite the vectors as
\be
{\bf \Psi}^{(j)}=\left|{\bf \Psi}^{(j)}\right|\,\left(\sin\theta_j\exp(\ui\phi_j),\cos\theta_j{\bf u}^{(j)}\right),
\ee
for real $\theta_j,\phi_j$, and where ${\bf u}^{(j)}$ are normalized complex vectors of dimensionality $\mu-1$. We can then write
\be
I_1\ge\left|{\bf \Psi}^{(0)}\right|^2\cos^2\theta_0-\left|{\bf \Psi}^{(1)}\right|^2\sin^2\theta_1.\label{ge1}
\ee

To conclude the proof, we use the fact that Eq.~(\ref{perpendicular}) implies that the vectors ${\bf \Psi}^{(j)}$ are orthogonal, {\it i.e.}, ${\bf \Psi}^{(1)*}\cdot{\bf \Psi}^{(0)}=0$, which can be written as
\be
\cos\theta_0\cos\theta_1{\bf u}^{(1)*}\cdot{\bf u}^{(0)}=-\sin\theta_0\sin\theta_1\exp[\ui(\phi_0-\phi_1)].
\ee
The square modulus of both sides of this expression together with the fact that $|{\bf u}^{(1)*}\cdot{\bf u}^{(0)}|\le1$ lead to 
\be
\cos^2\theta_0\cos^2\theta_1\ge\sin^2\theta_0\sin^2\theta_1,
\ee
which, by adding $\cos^2\theta_0\sin^2\theta_1$ to both sides, becomes
\be
\cos^2\theta_0\ge\sin^2\theta_1.\label{it}
\ee
This inequality, together with Rel.~(\ref{2ge1}), imply that the right-hand side of Rel.~(\ref{ge1}) is greater than or equal to zero, and therefore $I_1\ge0$.

Let us finish by noting that there are states that achieve the equality. For this to be true, two conditions must be satisfied. First, the inequality in Rel.~(\ref{it}) must be an equality, and this requires the two vectors ${\bf u}^{(j)}$ to be parallel. Second, the terms for $m>1$ within the sum in Eq.~(\ref{stepw}) must vanish, which happens if the vectors ${\bf u}^{(j)}$ are restricted to the $(N-1)$-dimensional subspace on which ${\bf w}$ has zero components. Since ${\bf u}^{(1)}\propto{\bf u}^{(2)}$, this is certainly possible for $N\ge1$.

\subsection{Derivation of entanglement inequalities for two qubits and a qu-tetrit}
\begin{figure}
\begin{center}
\includegraphics[scale=0.35]{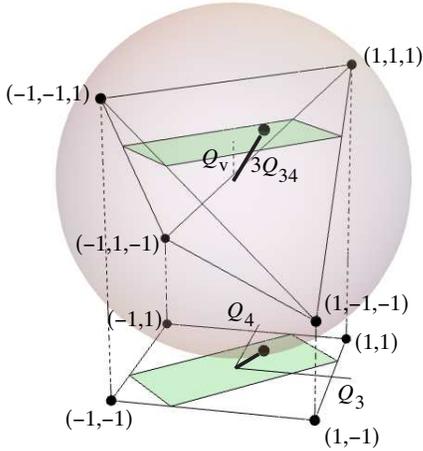}
\caption{Relation between the center-of-mass geometry in three dimensions for a party with four states, corresponding to four equidistant masses over the surface of a sphere, and the geometry in two dimensions for two parties with two states each, corresponding to four masses at the corners of a square, shown at the bottom. This relation consists of a vertical projection, following scaling the three-dimensional case by a factor of $\sqrt{3}$ (which is then the radius of the sphere).}\label{FigAppendix2}
\end{center}
\end{figure}

We now derive the constraints relating the entanglement measures $Y_1$ and $Y_2$ for two qubits and $Y_{34}$ for a qu-tetrit by using the center-of-mass geometry. 
The key is to find a connection between the geometries for one qu-tetrit party and for two qubit parties. Since party 34 has four states, the subspace leading to the geometric interpretation of $Q_{34}$ is three-dimensional, where $Q_{34}$ is the magnitude of the center of mass vector for four point masses at the four corners of a regular tetrahedron inscribed in a unit sphere, say, at the points $(1,1,1)/\sqrt{3}$, $(-1,-1,1)/\sqrt{3}$, $(-1,1,-1)/\sqrt{3}$, $(1,-1,-1)/\sqrt{3}$. On the other hand, for the separate parties 3 and 4, the subspace leading to the geometric interpretation of both $Q_3$ and $Q_4$ is two-dimensional, these quantities corresponding to the two coordinates of the centers of mass of four point masses at the corners of a square of side 2, that is, at the points $(1,1)$, $(1,-1)$, $(-1,-1)$, $(-1,1)$. The geometric connection between these two geometries with different dimensionality results from noting that the projection over the third coordinate of the four corners of the tetrahedron onto two dimensions results in a square of side $2/\sqrt{3}$. By multiplying by a scaling factor of $\sqrt{3}$, this square is made to correspond to a square of side 2 centered at the origin with the four point masses at its corners. This geometric correspondence is shown in Fig.~\ref{FigAppendix2}.

As can be seen in Fig.~\ref{FigAppendix2}, the following relation holds
\be
3Q_{34}^2=Q_3^2+Q_4^2+Q_{\rm v}^2,\label{threecoords}
\ee 
where $Q_{\rm v}$ is the third coordinate of the center of mass in the three-dimensional space.  Further, since the center of mass must be inside the tetrahedron, for any given $Q_{\rm v}$, the cross-section of the allowed region for $Q_3$ and $Q_4$ is a rectangle along the bisector of the $Q_3$ and $Q_4$ axes, shown in green in the figure, and defined by the restrictions (where we assume $Q_3,Q_4,Q_{\rm v}\ge0$):
\bse
\bea
(Q_3+Q_4)\le1+Q_{\rm v},\label{Q3Q4bound}\\
|Q_3-Q_4|\le1-Q_{\rm v}.
\eea
\ese

The desired relation is found by starting from the inequality for the four qubit system:
\be
|Y_1-Y_2|\le Y_3+Y_4=2-(Q_3+Q_4).
\ee
For $Y_3+Y_4>1$ this inequality is trivial, so we only pay attention to the case $Y_3+Y_4\le1$ or, equivalently, $Q_3+Q_4\ge1$. The goal is to write this expression in terms of $Y_{34}=1- Q_{34}$. That is, we must find the smallest value of $Q_3+Q_4$ for given $Q_{34}$. Fixing $Q_{34}$ restricts the center of mass in the space $(Q_3,Q_4,Q_{\rm v})$ to the surface of a sphere of radius $\sqrt{3}Q_{34}$, and for $\sqrt{3}Q_{34}\ge\sqrt{2}$, the smallest value of $Q_3+Q_4$ (which is still larger than unity) occurs when both $Q_{\rm v}$ and either $Q_3$ or $Q_4$ equal unity. Letting, for example, $Q_4=1$ implies that party 4 is not entangled, so we can use instead the relation for three qubits $|Y_1-Y_2|\le Y_3$. Since $Y_3=1-\sqrt{2{\rm tr}\rho_3^2-1}$ and ${\rm tr}\rho_3^2={\rm tr}\rho_{34}^2=\sqrt{3(1-Y_{34})^2+1}/2$ (assuming $Y_4=0$), this inequality yields
\be
|Y_1-Y_2|\le1-\sqrt{\frac{H[3(1-Y_{34})^2-1]}2}.
\ee

To derive the second relation, we simply use relation~(\ref{Q3Q4bound}) to eliminate $Q_{\rm v}$ from Eq.~(\ref{threecoords}), to find
\be
3Q_{34}^2\ge Q_3^2+Q_4^2+H^2(Q_3+Q_4-1).
\ee
Notice that if we were to combine parties 1 and 2 into a four-state ``party 12'', the equality $Q_{12}=Q_{34}$ would hold. Therefore, we can replace $Q_{34}$ with $Q_{12}$ in the relation above, or equivalently, replace $Q_3$ and $Q_4$ with $Q_1$ and $Q_2$. This second substitution leads to 
\be
Y_{34}\le1 - \sqrt{\frac{(1 - Y_1)^2 + (1 - Y_2)^2 + H^2(1 - Y_1 - Y_2)}3}
\ee

\subsection{Relations for shared classical resources}
As mentioned in the manuscript, the polygon inequalities relating the measures $Y_n$ are fully analogous to those that apply to shared classical resources, such as real estate. Consider $N$ property owners, each of whom owns the same amount of property (normalized to unity for simplicity). The total property is divided in parts labeled as $P_n$ that are owned by an individual party, and parts labeled as $P_{nm}$ that are co-owned in equal amounts by two parties. For example, in the case of three parties, the total amounts of property of each party are
\bse
\bea
{\rm party 1:}\,\,\,\,1&=&P_1+\frac{P_{12}+P_{13}}2,\\
{\rm party 2:}\,\,\,\,1&=&P_2+\frac{P_{12}+P_{23}}2,\\
{\rm party 3:}\,\,\,\,1&=&P_3+\frac{P_{13}+P_{23}}2.
\eea
\ese
(Notice that there is no need to add property contributions co-owned by more than two parties, say $P_{123}$, since they could be separated into equal parts, each co-owned by a different pair of parties, without changing the results that follow.) The measures of  ``ownership entanglement'' correspond to the fraction of each party's property that is co-owned:
\bse
\bea
Y_1&=&\frac{P_{12}+P_{13}}2,\\
Y_2&=&\frac{P_{12}+P_{23}}2,\\
Y_3&=&\frac{P_{13}+P_{23}}2.
\eea\label{classicY}
\ese
Perhaps abusing the quantum analogy, one can justify referring to these quantities as entanglement measures since any party $n$ is unaffected by processes (say, property redistributions) that apply to all other parties only if $Y_n=0$, and is maximally affected if $Y_n=1$. 

The polygon inequalities for this situation are straightforwardly verified. For example, from Eqs.~(\ref{classicY}) we find $Y_2+Y_3-Y_1=P_{23}\ge0$, the equality holding only if parties 2 and 3 do not share any property. In fact, given this relation, the combination $Y_2+Y_3-Y_1$ isolates the entanglement exclusively between parties 1 and 2. 

\begin{figure}
\begin{center}
\includegraphics[scale=0.5]{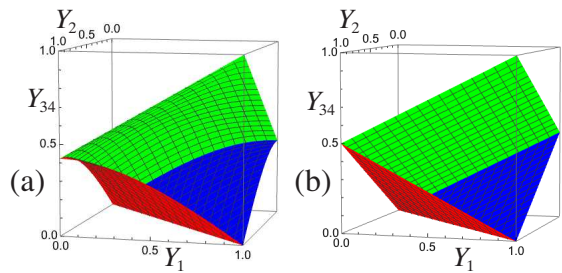}
\caption{Boundaries between the allowed and forbidden regions for $Y_1$, $Y_2$, and $Y_{34}$ for (a) the quantum entanglement relations derived in the previous section of these supplementary materials, and (b) the shared classical resource relations (\ref{classicrelation}). In both cases, the allowed regions are those on the side of the concavity formed by the three surfaces.}
\label{FigAppendix3}
\end{center}
\end{figure}

The polygon inequalities for a classical resource shared amongst $N$ parties discussed above are identical to those for the quantum entanglement measures $Y_n$ for a $N$-party pure state. However, for $N\ge4$, it is possible to consider also bipartitions including more than one party on each side. For those, the analogy between quantum qubits considered in the previous section and classical property owners is not completely accurate, as we now show. Consider the case of $N=4$. As in the four-qubit case described in the manuscript, let parties 3 and 4 merge into a ``party 34''. The property owned by this party would be
\bea
{\rm party 34:}\,\,\,\,2&=&P_3+P_4+P_{34}\nonumber\\
&+&\frac{P_{13}+P_{23}+P_{14}+P_{24}}2
\eea
so that the fraction of this party's property shared with the other parties is
\be
Y_{34}=\frac{P_{13}+P_{23}+P_{14}+P_{24}}4.
\ee
On the other hand
\bse
\bea
Y_1&=&\frac{P_{12}+P_{13}+P_{14}}2,\\
Y_2&=&\frac{P_{12}+P_{23}+P_{24}}2,
\eea
\ese 
from where we can find
\be
Y_1+Y_2=P_{12}+2Y_{34}.\label{P12rel}
\ee
Since $0\le P_{12}\le 2\,{\rm min}(Y_1,Y_2)$, we find the inequalities
\be
\frac{|Y_1-Y_2|}2\le Y_{34}\le\frac{Y_1+Y_2}2.\label{classicrelation}
\ee
Of course, similar inequalities would apply to all other bipartitions separating pairs of parties. Note also that, as in the quantum case, $Y_{12}=Y_{34}$. Figure~\ref{FigAppendix3} shows a comparison between the restrictions to $Y_{34}$ in terms of $Y_1$ and $Y_2$ for the qubit entanglement case (also shown in the main body of the article), as well as for the classical resource sharing case in (\ref{classicrelation}). Note that the two allowed regions agree largely but not entirely, the quantum restriction having curved boundaries in contrast with the flat ones of the classical restriction. For the quantum restriction, the volume of the allowed region is $0.3457$ while for the classical restriction it is $1/3$, and the volume of their intersection is $0.3024$. 

Finally, notice from Eq.~(\ref{P12rel}) together with the fact that $Y_{12}=Y_{34}$ that we can define a measure of classical resource sharing between any two parties $n$ and $m$ as $p_{nm}=P_{nm}/2=(Y_n+Y_m)/2-Y_{nm}$, valid for any $N$ and that takes values between zero (the two parties are unentangled) and unity (the states are fully entangled). This measure is valid only if both parties own the same total amount of the classical resource in question. Correspondingly, if this measure were to be used to characterize the quantum entanglement between two parties, these parties must have the same number of states. This quantity has some problems in the limit of small entanglement between the parties when the density matrix $\rho_{nm}$ is significantly mixed. 

\subsection{Center-of-mass-based proof of entanglement inequalities for three-qubit states}
The interpretation of ${\bf Q}$ as a center of mass allows also an alternative, geometric proof for the inequalities in Rel.~(\ref{ineq}), at least for $N=3$. Note that, in the geometric construction, the points corresponding to the lowercase components, represented by blue spheres in Fig.~4 of the main manuscript, are at the four corners of a regular tetrahedron, and the same with the uppercase ones, represented by green spheres in this figure. Also, the point with mass $|a|^2$ is opposite that with mass $|A|^2$, and so on. 
The key to the proof are the Schmidt representation constraints in Eq.~(6) of the main manuscript, which when using the notation introduced in the main manuscript's Eq.~(12), can be written as
\begin{subequations}\label{conditionslines}
\bea
aB^*+Cd^*+Dc^*+bA^*&=&0,\\
aC^*+Db^*+Bd^*+cA^*&=&0,\\
aD^*+Cb^*+Bc^*+dA^*&=&0.
\eea
\end{subequations}

Proving relation (11) in the main manuscript is equivalent to proving that the center of mass ${\bf Q}$ must be located within either or both of the two tetrahedra whose corners are the corners of the cube, since the union of these two volumes is precisely the region allowed by the main manuscript's relation (11). To show that ${\bf Q}$ must be inside at least one of these tetrahedra, we write it as
${\bf Q}=g{\bf v}+G{\bf V}$, 
where ${\bf v}$ and ${\bf V}$ are the centers of mass for the masses corresponding to lowercase and uppercase coefficients:
\begin{subequations}\label{vectorsvV}
\bea
{\bf v}&=&(|a|^2+|b|^2-|c|^2-|d|^2,\nonumber\\&&\,\,|a|^2-|b|^2-|c|^2+|d|^2,\nonumber\\&&\,\,|a|^2-|b|^2+|c|^2-|d|^2)/g,\\
{\bf V}&=&(-|A|^2-|B|^2+|C|^2+|D|^2,\nonumber\\&&\,\,-|A|^2+|B|^2+|C|^2-|D|^2,\nonumber\\&&\,\,-|A|^2+|B|^2-|C|^2+|D|^2)/G,
\eea
\end{subequations}
with $g=|a|^2+|b|^2+|c|^2+|d|^2$ and $G=|A|^2+|B|^2+|C|^2+|D|^2$ being the total masses for each subset, so that $g+G=1$. 
It is shown in what follows that ${\bf v}$ and ${\bf V}$ are parallel, that is, collinear with the origin. Since the global center of mass ${\bf Q}$ must be along the straight line segment joining the partial centers of mass ${\bf v}$ and ${\bf V}$, it also must be inside at least one of the two tetrahedra and therefore in the region allowed by (11).

To show the proportionality of ${\bf v}$ and ${\bf V}$, we use Eqs.~(\ref{conditionslines}). The first two of these equations can be rearranged as
\bea
|aB^*+Cd^*|^2&=&|Dc^*+bA^*|^2,\\
|aC^*+Bd^*|^2&=&|Db^*+cA^*|^2.
\eea
After expanding each side, the difference of these two relations can be written as
\be
(|a|^2-|d|^2)(|B|^2-|C|^2)=(|A|^2-|D|^2)(|b|^2-|c|^2).
\ee
Two more such equations can be derived by using different pairs of Eqs.~(\ref{conditionslines}). Collectively, the three resulting equalities can be written as
\bea
\frac{|A|^2-|B|^2}{|a|^2-|b|^2}&=&\frac{|C|^2-|D|^2}{|c|^2-|d|^2}=\kappa_1,\\
\frac{|A|^2-|C|^2}{|a|^2-|c|^2}&=&\frac{|B|^2-|D|^2}{|b|^2-|d|^2}=\kappa_2,\\
\frac{|A|^2-|D|^2}{|a|^2-|d|^2}&=&\frac{|B|^2-|C|^2}{|b|^2-|c|^2}=\kappa_3.
\eea
We now show that the three ratios $\kappa_n$ are actually equal. Consider, for example, rewriting the first and third relations in the forms
\bea
\frac{|a|^2-|b|^2}{|c|^2-|d|^2}+1&=&\frac{|A|^2-|B|^2}{|C|^2-|D|^2}+1,\\
\frac{|a|^2-|d|^2}{|b|^2-|c|^2}-1&=&\frac{|A|^2-|D|^2}{|B|^2-|C|^2}-1.
\eea
After taking common denominators and using the third component of Eqs.~(\ref{vectorsvV}), one can rewrite these relations as
\bea
\frac{gv_3}{|c|^2-|d|^2}&=&\frac{GV_3}{|C|^2-|D|^2},\\
\frac{gv_3}{|b|^2-|c|^2}&=&\frac{GV_3}{|B|^2-|C|^2},
\eea
so that
\be
GV_3=\kappa_1gv_3=\kappa_3gv_3.
\ee
and therefore $\kappa_1=\kappa_3$. One can similarly show that $\kappa_2=\kappa_1$, and that
\be
G{\bf V}=\kappa_1g{\bf v},
\ee
hence proving the proportionality of the two vectors.

\end{document}